\documentclass[5p, times]{elsarticle}

\usepackage{datetime}
\usepackage{amsmath}
\usepackage{amsfonts}
\usepackage{graphicx}
\usepackage{amssymb}
\usepackage{amssymb}
\usepackage{ latexsym }
\usepackage{hyperref}
\usepackage{xcolor}
\usepackage{graphicx}
\usepackage{dcolumn}
\usepackage{bm}

\usepackage{enumitem}
\usepackage{datetime}
\usepackage{amsmath}
\usepackage{amsfonts}
\usepackage{graphicx}
\usepackage{amssymb}
\usepackage{amssymb}
\usepackage{latexsym}
\usepackage{hyperref}
\usepackage{array}
\usepackage{bbold}

\newcommand\figurewidth{1\linewidth}

\def\abs#1{\left\lvert#1\right\rvert}

\newcommand{\sign}{{\mathop{\mathrm{sign}}}}

\usepackage{xcolor}
\definecolor{mygreen}{HTML}{006E28}

\begin{document}
\begin{frontmatter}
\title{Staccato radiation from the decay of large amplitude oscillons}
\begin{abstract}
We study the decay of large amplitude, almost periodic breather-like states in a deformed sine-Gordon model
in one spatial dimension.
We discover that these objects decay in a staggered fashion via a series of transitions, during which higher harmonics are released as short, staccato bursts of radiation.
Further, we argue that this phenomenon
is not restricted to one particular model, and that
similar mechanisms of radiative decay of long-lived oscillating
states can be observed for a wide class of physical systems,
including the $\phi^6$ model.
\end{abstract}
\author{Patrick Dorey}
\ead{p.e.dorey@durham.ac.uk}
\address{Department of Mathematical Sciences, Durham University, UK}
\author{Tomasz Roma\'nczukiewicz}
\ead{trom@th.if.uj.edu.pl}
\address{Institute of Physics, Jagiellonian University, Krak\'ow, Poland}
\author{Yakov Shnir}
\ead{shnir@maths.tcd.ie}
\address{BLTP, JINR, Dubna 141980, Moscow Region, Russia}
\address{Department of Theoretical Physics, Tomsk State Pedagogical University, Russia}
\begin{keyword}
oscillons, relaxation, nearly-integrability
\end{keyword}

\end{frontmatter}

Many non-linear physical systems support
oscillons,  spatially localized almost time-periodic field configurations
which can live for exceptionally long times \cite{bogolyubsky1976pulsed,copeland1995oscillons,gleiser1994pseudostable}.
Unlike the breathers
of the sine-Gordon (sG) model, oscillons (sometimes also referred to as quasi-breathers) continuously emit scalar radiation
via the excitation of scattering modes of the continuous spectrum.
Oscillons 
and their close relatives, gravitationally bound oscillatons,
appear in a broad class of classical
field theories in various dimensions,
and in recent
years they 
have attracted much attention
\cite{Riotto:1995yy, Umurhan:1998ez, Farhi:2005rz, Gleiser:2006te, Alcubierre:2003sx,Graham:2006vy,Fodor:2009kg}.

In 1+1 dimensional scalar field theories, the slow decrease of the
amplitude
of small oscillons continues
smoothly up to the limit of very small size. However, and despite the
fact that its amplitude
becomes arbitrarily small, the oscillon can never be described as a solution of a linearized
equation in the vacuum sector, since its existence depends on the
nonlinearity of the system. Sine-Gordon breathers possess exactly the
same feature, and
the similarity between the perfectly periodic, non-radiating
sG breathers and oscillon solutions of the $\phi^4$ theory
leads to interesting relationships between the two,
something that we will exploit in this paper.
It is worth mentioning that modified sG models, and variants including
the  $\phi^4$ and $\phi^6$ theories, have
many physical applications and have been studied for many decades
\cite{PhysRevLett.102.224101, RevModPhys.61.763, Cuevas–Maraver2019,
sgBook2014, Gulevich, FLACH1998181, PhysRevD.12.1606, Lohe:1979mh}.
Nevertheless, 
we were able to find the novel phenomenon of stacatto decay, as
described in this paper.

The decay rate of
small-amplitude oscillons in nonintegrable models such as the $\phi^4$
model has been explored in a number of works, including 
\cite{segur1987nonexistence, BOYD1995311, fodor2009computation, Fodor:2019ftc}.
It was found to be beyond all orders in perturbation
theory,
given by $dE/dt ~\sim -\exp{(-B/E)} $, where $E$ is the energy of the
oscillon and $B$ is a constant.
The long time evolution of the small-amplitude
oscillon thus involves the slow decrease of its amplitude, which turns out to be accompanied by a gradual
increase in its frequency. Nevertheless, the frequency remains  below the mass threshold, meaning that the modes
of the continuum can only be excited through the second (and higher) harmonics
\cite{fodor2009computation,hindmarsh2008oscillons,honda2002fine}. In higher spatial dimensions, the configuration
may rapidly collapse into radiation as the fundamental frequency approaches some critical value, which is still
below the mass threshold \cite{fodor2009computation,hindmarsh2008oscillons, honda2002fine, Hindmarsh2012, PhysRevLett.101.011602, PhysRevD.80.125037}.
However, and despite many extensive numerical studies, there is still very
little known about the true nature of this decay process.

In contrast to previous studies, our focus here is not on small amplitude
oscillons,
but rather on large oscillons, taking a mildly deformed 1+1 dimensional sG
model with weak integrability breaking as our main example.
This deformation lifts the infinite degeneracy of the vacuum while leaving
a $\mathbb{Z}_2$ symmetry unbroken. We have found that the long time
evolution of the corresponding oscillon configuration is not smooth,
but is realised through a series of transitions caused by the release
of higher propagating harmonics.
In previous work \cite{Dorey:2017bdr, OscillonsExternal2018} we
found that deep, oscillating bound modes, when excited to the
nonlinear regime, can, in certain circumstances, block one or more
harmonics as possible decay channels. This phenomenon depends on the nonlinear
dependence between frequency and amplitude, which in the previously studied
models could be found only numerically, or  within perturbation
theory.  Breathers, on the other hand, are perfect examples of such
nonlinear excitations, and for suitably small deviations from
integrability they well approximate the decaying oscillon.
Their frequency-amplitude relation is well
known analytically.  Moreover, the frequency can be changed, due to
the nonlinearities, within the entire mass gap.  By breaking the
integrability we allow the oscillons to radiate and evolve
through a wide range of frequencies.

\noindent
\textit{~~A deformed sine-Gordon model.~}
Consider a scalar field theory in 1+1 dimensions defined by the Lagrangian
\begin{equation}
\mathcal{L}=\frac{1}{2}\phi_t^2-\frac{1}{2}\phi_x^2-U(\phi) \,,
\end{equation}
where $U(\phi)$ is a simple
$\mathbb{Z}$-breaking
modification of the sG potential
depending on a parameter $\epsilon
\in [0,1]$:
\begin{equation}
 U(\phi) = (1-\epsilon)(1-\cos\phi)+\frac{\epsilon\phi^2}{8\pi^2}(\phi-2\pi)^2\,.
 \label{potential_1}
\end{equation}
The standard sG
potential is recovered in the limit $\epsilon=0$, while
setting $\epsilon=1$ yields a potential
with just two vacua, $\phi_{v_1}=0$ and $\phi_{v_2}=2\pi$,
a shift and rescaling of the
$\mathbb{Z}_2$-symmetric potential of the $\phi^4$ model.
Thus, the model with
the potential (\ref{potential_1}) interpolates between the
integrable sG model and the $\phi^4$ model, with integrability
being broken
for all non-zero values of $\epsilon$. Note that the
parameters in (\ref{potential_1}) are fixed in such a way
that the mass of small perturbations around the vacuum remains the same as
in the original sG model, since $U''(0)=m^2=1$ for all values of
$\epsilon$.

The sG model supports a spatially localized solution, exactly
periodic in time, the breather:
\begin{equation}
 \phi_B(x, t;\omega, b)=4\,\arctan\left(\frac{b\cos\omega t}{\sqrt{1-b^2}\cosh bx}\right)\,,
 \label{breather}
\end{equation}
where $b=\sqrt{1-\omega^2}$. The topological charge of this
configuration is zero, and it can be viewed as a bound state of a kink and anti-kink,
oscillating with a constant frequency $\omega$, which is
a parameter of the solution.  Although $b$ is a function of the
frequency $\omega$, we leave the dependence on $b$ explicit to
distinguish which part of the solution is associated with the
profile and which with the time evolution.

Our goal is to study the time
evolution of the breather configuration (\ref{breather}) in the
deformed model (\ref{potential_1}).

Note that in the pure sG model, the breather solution can be decomposed into a Fourier series in time
with only odd multiplicities of the frequency $\omega$.
Since the deformed potential (\ref{potential_1}) is no longer
invariant with respect to parity reflections about either vacuum,
the corresponding expansion for
$\epsilon\neq 0$ should also include even harmonics.

The oscillon solutions of the $\phi^4$ theory have been studied in
many papers, including
\cite{segur1987nonexistence,BOYD1995311,
fodor2009computation,hindmarsh2008oscillons,honda2002fine}.
When the amplitude of the oscillations is small, a $\phi^4$
oscillon can be treated as a small sG breather perturbed by higher
order polynomial terms in the
potential. Evidently, this deformation breaks the integrability of the
sG system and causes
the radiative decay of the oscillon.

In this paper we consider a different limit. We assume that the
amplitude is large, but the modification of the sG potential
is small, $\epsilon\ll 1$.  We conjecture that in such a case the
profile of the oscillon can be well approximated by the sG breather,
thus the initial data in our  numerical simulation corresponds to
the configuration (\ref{breather})
\begin{equation}
 \phi(x,0)=\phi_B(x,0;\omega_0,b_0)\,,\qquad \phi_t(x,0)=0
\end{equation}
for some initial frequency $\omega_0$, with
$b_0=\sqrt{1-\omega_0^2}$\,.

\begin{figure}[ht]
 \centering
 \includegraphics[width=\linewidth]{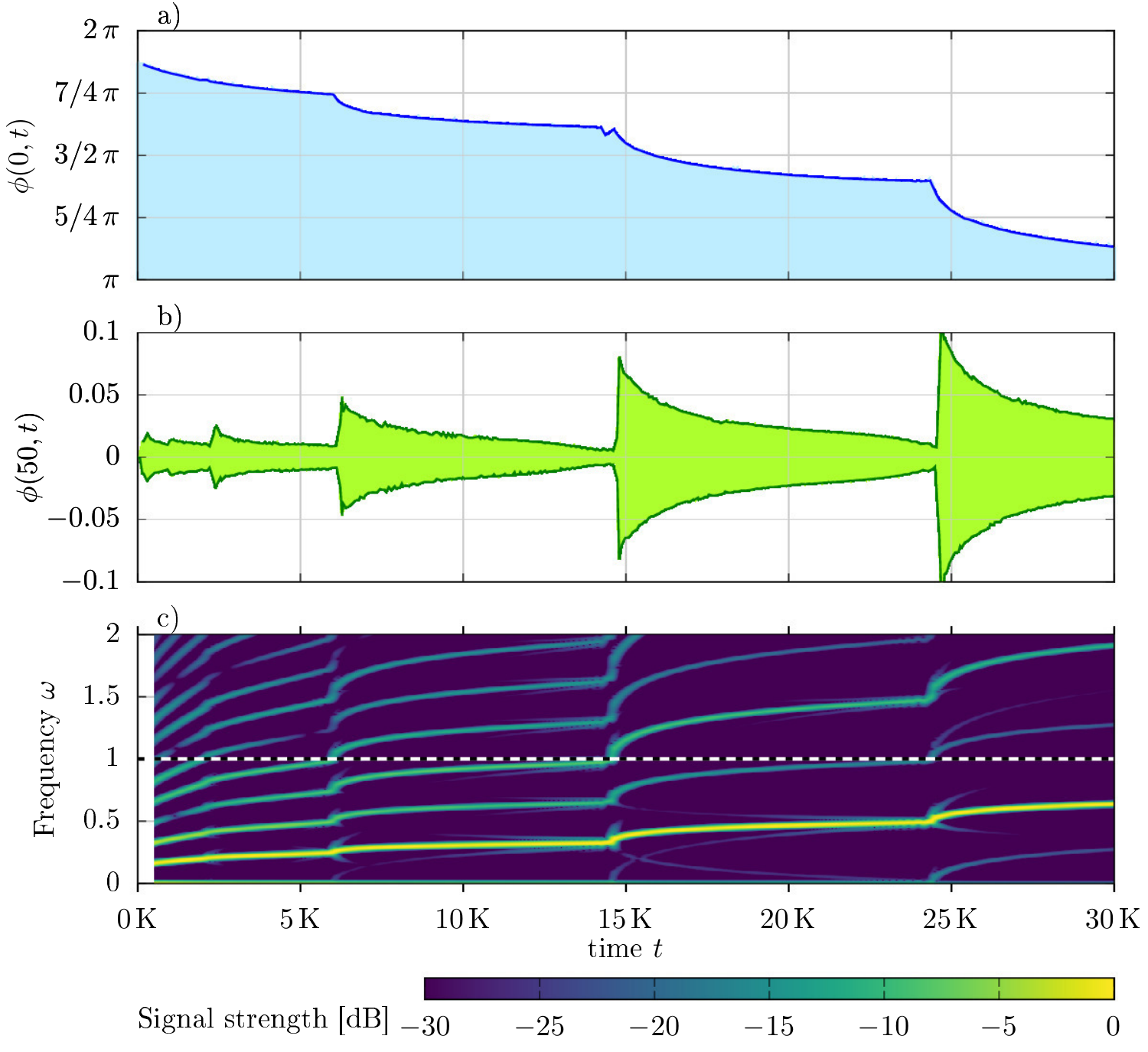}
 \caption{Multiple transitions for $\epsilon=0.0025$, $\omega_0=0.1$
at frequencies $\omega=m/n$, $n=2\ldots6$. The mass threshold $m=1$ is
denoted by a dashed black and white line.
(a) Amplitude of the oscillations at the centre $x=0$,
(b) radiation amplitude measured at $x=50$,
(c) spectrogram of the field measured at the centre  $\phi(0,t)$.
}\label{ExampleDecay1}
\end{figure}

\noindent
\textit{~~Numerical simulations.~}
We have solved the full nonlinear evolution partial differential
equation numerically, taking the
breather profile (\ref{breather}) as the initial condition at $t=0$.
Large amplitude oscillons with $\omega<\frac{1}{2}$ radiated visibly
even for relatively small values of the integrability breaking
parameter, $\epsilon<0.001$.
We measured the field values at the centre $x=0$, and far away from
the oscillon, $x=50$. An example evolution for $\omega_0=0.1$ and
$\epsilon=0.0025$ is
shown in the Figure \ref{ExampleDecay1} 
(see -- and hear -- also the Supplementary Material \ref{av:A} and \ref{av:B}). 
Subplot (a) shows the field at the centre, (b) the field at
a distance $x=50$ from the centre and (c) the spectrogram of the field measured at the centre.

The most striking feature of the evolution is that the oscillon does
not relax uniformly (Figure \ref{ExampleDecay1}(a)). Relatively long
times of what looks like the standard relaxation processes are
separated by sudden sharp jumps, during which the rate of the decay
increases significantly. These jumps are reflected in far-field
measurements (Figure \ref{ExampleDecay1}(b)) as larger
bursts of radiation, by at least one order of magnitude.
The spectrogram (c) gives more insight into the nature of these jumps.
Bright lines correspond to dominant frequencies in the spectrum.
Notably, the lowest, \textit{basic}, frequency is accompanied by all
of its multiples.
After a short transient time of some adjustment, at least six harmonics are visible below the mass threshold.
For most of the time the basic frequency and the higher
harmonics change adiabatically slowly. A jump occurs whenever one of
the harmonics crosses the value $\omega=1$, which is the mass threshold.
The frequency of the oscillon quickly changes and then slows down until the next event.
During these short transitions the harmonic entering the scattering spectrum changes its nature from localized to wave-like. This explains the burst of radiation observed in the far field.

\noindent
\textit{~~Stationary approach.~}
Numerical simulations suggest that the oscillon for most of the time
evolves adiabatically, slowly changing its basic frequency and amplitude. It is therefore justified to assume that, in between the jumps, the momentary state of an oscillon can be well approximated by a stationary solution, which is just small perturbation of a breather. Let us consider a perturbative expansion of the field equation
\begin{equation}\label{PDEEq}
 \phi_{tt}-\phi_{xx}+\sin\phi+\epsilon \delta U'(\phi)=0 \ ,
\end{equation}
around the breather solution (\ref{breather}),
$\phi=\phi_B+\epsilon\phi^{(1)}+\mathcal{O}(\epsilon^2)$.
The deformation of the usual sG potential defined by
(\ref{potential_1}) is of the first order in $\epsilon$,
\begin{equation}
 \delta U=\frac{\phi^2}{8\pi^2}(\phi-2\pi)^2-1+\cos\phi\, .
 \label{pert}
\end{equation}
In the first order in $\epsilon$ the corresponding equation of motion is
\begin{equation}\label{linearsG}
 \phi^{(1)}_{tt}-\phi^{(1)}_{xx}+\left(\cos\phi_B\right)\phi^{(1)}+\delta U'(\phi_B)=0\,.
\end{equation}
Here
\begin{equation}
 \cos\phi_B=\frac{1-6u^2+u^4}{(1+u^2)^2}\,,\qquad
 u = \frac{\sqrt{1-\omega^2}\;\cos(\omega t)}{\omega\;\cosh(\sqrt{1-\omega^2}\; x)} \, .
\end{equation}

Note that the partial differential equation (\ref{linearsG}) resembles
the  well known Mathieu or Hill equation with additional driving force $\delta U'$.

The breather solution is periodic in time, and the spectrogram of the
oscillon field at its centre shows that it is almost periodic, with a
well defined basic frequency and its multiples.
Therefore we can make use of the Fourier decomposition
\begin{equation}
 \phi^{(1)}=\sum_{n=-\infty}^\infty\xi_n(x)e^{in \omega t}\,,\quad\cos\phi_B=\sum_{n=-\infty}^\infty V_n(x) e^{in \omega t}
\label{breather-expansion}
\end{equation}
where $V_{-n}=V_n$, $\xi_{-n}=\xi_n^*$ and
\begin{equation}
  \delta U'(\phi_B)=\sum_{n=-\infty}^\infty g_n(x)e^{ in \omega t}\,.
\end{equation}
Substituting these expansions into the first order equation (\ref{linearsG}), we obtain
\begin{equation}\label{ODESys}
 \left(-n^2\omega^2-\frac{d^2}{dx^2}\right)\xi_n+\sum_{m=-\infty}^\infty V_{n-m}\xi_m+g_n=0\,.
\end{equation}
This is a system of linear, coupled ODEs, which describes the evolution
of the modes $\xi_n$, coupled via the convolution of infinite vectors $\xi_m$ and $V_n$.
The solutions to equation (\ref{ODESys}) for different harmonics $\xi_n$ can be categorised as either
normalizable modes with $|n\omega|<1$, which are localized by the
potential $V_n(x)$, or radiative modes
above the mass threshold, $|n\omega|>1$.
The latter modes represent outgoing waves, and satisfy the following boundary condition for $L\gg 1$
\begin{equation}
 i k_n\xi_n(\pm L)\pm\xi_n'(\pm L)=0,\quad k_n=\sign(n)\sqrt{\abs{1-n^2\omega^2}}\, ,
\end{equation}
which ensures the absence of the incoming waves.

The normalizable Fourier modes ($|n\omega|<1$) must vanish exponentially at spatial infinity as $\sim e^{-k_n\abs{x}}$.
The asymptotic boundary condition for these modes is
\begin{equation}\label{ODEbndrycnd}
 k_n\xi_n(\pm L)\pm\xi_n'(\pm L)=0\,,\qquad L\gg 1\,.
\end{equation}

\begin{figure*}[ht]
 \centering
 \includegraphics[width=0.75\linewidth]{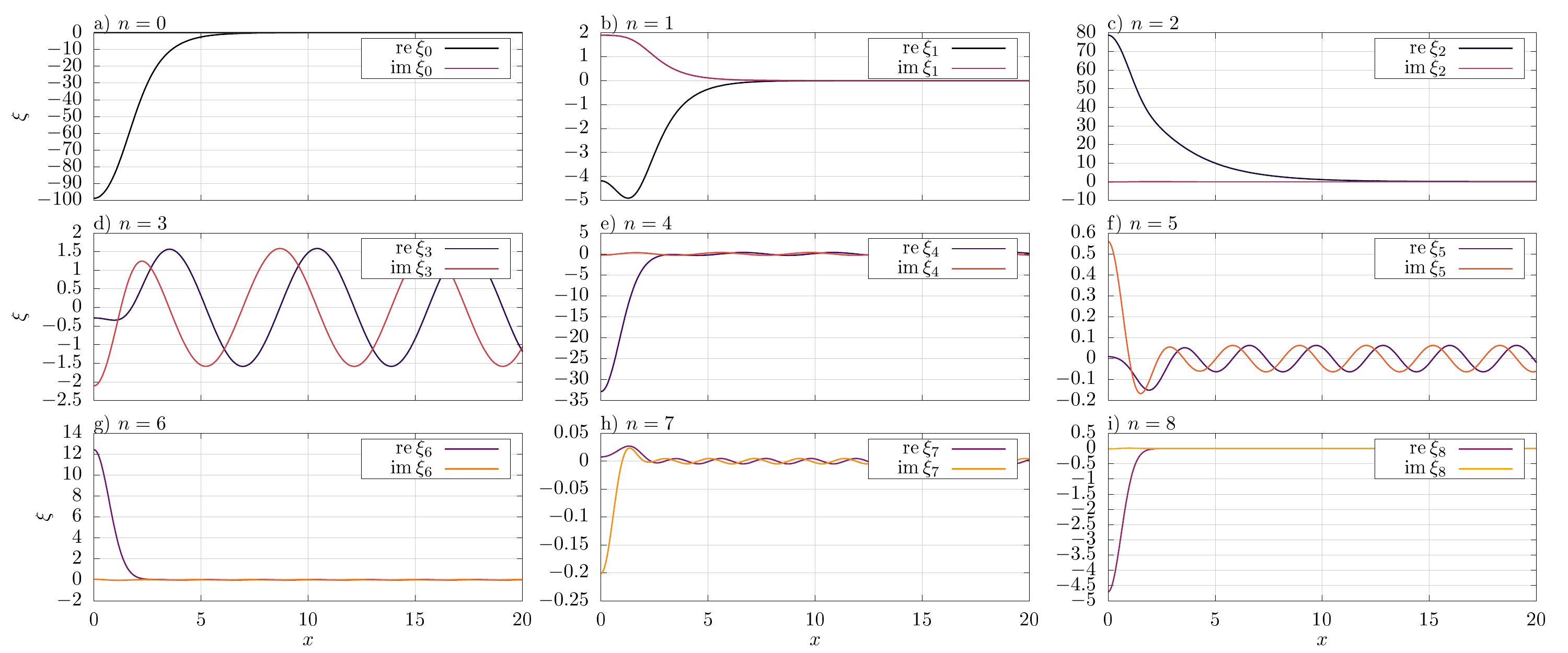}
 \caption{Example of the numerical solutions of the system of ODEs (\ref{ODESys}) for $\omega=0.45$.
 }\label{odeSolution}
\end{figure*}

We solved the above problem by
discretising the system (\ref{ODESys}) and
making use of the fourth order finite difference scheme with even boundary
conditions imposed at $x=0$. Example solution is plotted in Firgure \ref{odeSolution}.

From these solutions we can extract the amplitudes of the
corresponding modes $A_n$, defined as
\begin{equation}
 \xi_n(x\to L)\approx
 \begin{cases}
  A_n e^{-ik_nx}\,,&\text{for}\,|n|<1/\omega \\
  A_n e^{-k_nx}\,,&\text{for}\,|n|>1/\omega\,.
 \end{cases}
\end{equation}

\begin{figure}[ht]
 \centering
 \includegraphics[width=\columnwidth]{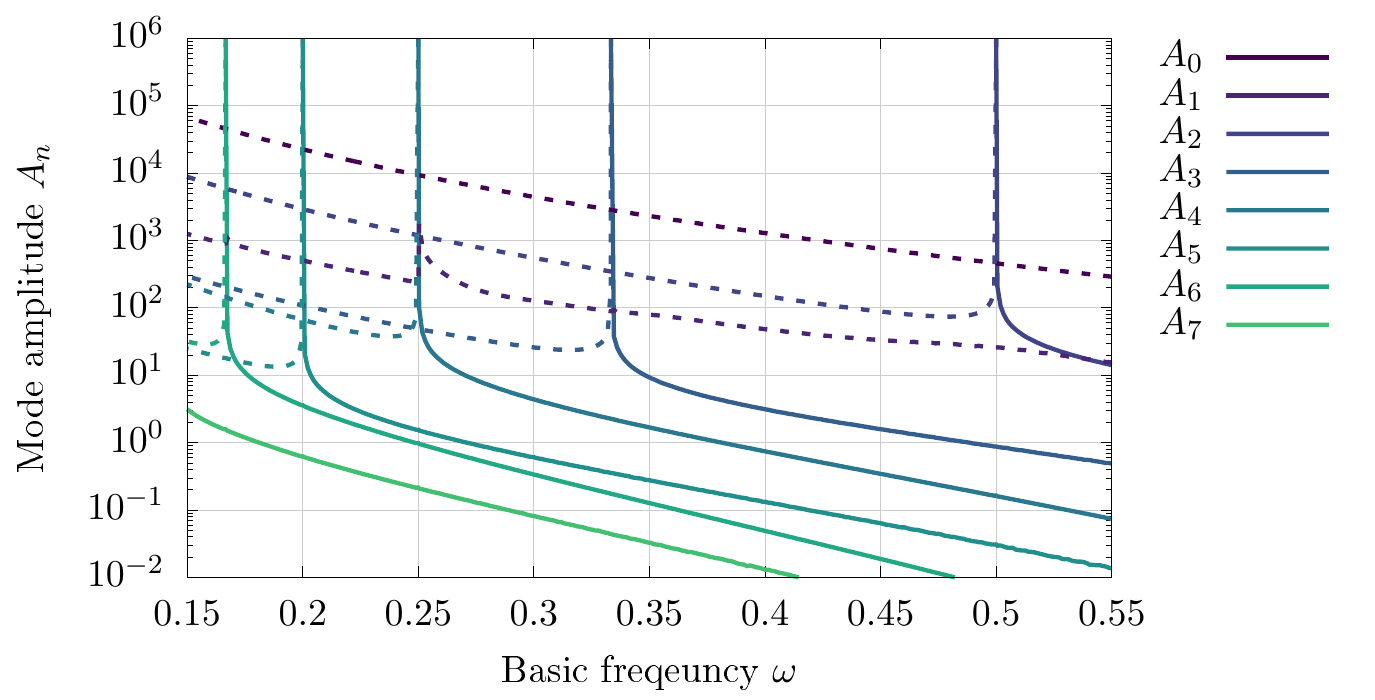}
 \caption{Mode amplitudes from the solution of the system of ODEs (\ref{linearsG}) for the first
 seven harmonics. Dashed lines correspond to the
non-propagating localized modes. Clearly visible peaks in the amplitudes correspond to the resonance frequencies
$\frac12, \frac13, \frac14, \frac15$ and $\frac16$.}\label{radiationAmplitudes}
\end{figure}

The amplitudes for the first few harmonics are shown in
Figure \ref{radiationAmplitudes}, where the Fourier series were truncated to include forty modes.
Note that one can clearly see the large amplitude peaks of the localized modes
at the resonance frequencies $\omega=1/n$,  $n=1,2 \dots$. These peaks correspond to the situation when one of the multiplicities of the basic frequency changes its nature from localized $n\omega<1$ to propagating $n\omega>1$. Exactly at $n\omega=1$ the wave number vanishes, $k_n=0$. The stationary approach described by equation (\ref{ODESys}) is unable to describe this situation properly due to the resonance between one of the harmonics and the mass threshold. This singularity shows the limitations of the perturbative approach and indicates that for these special cases the evolution is highly non-perturbative.
However, in between those resonant frequencies $\omega=1/n$, the evolution should be properly described by the Fourier decomposition, at least in the adiabatic approximation assuming that the frequency changes very slowly with time, which is true for small values of $\epsilon$. On the other hand, for all nonvanishing $\epsilon$ the oscillon should undergo a dramatic, non-perturbative  change in its structure (one of the harmonics delocalizes) in a close vicinity of the resonant frequencies.

One of the interesting features of the radiation bursts is their
characteristic structure, which allows the harmonic from which
they came to be identified. For example if the ratio of the first two observed
frequencies  is $3:2$ (a perfect fifth in musical terminology),
the second harmonic of the oscillon was released. If the ratio is
$4:3$ (a perfect fourth), then the third harmonic was released
and the second is still below the threshold 
(see and hear also the Supplementary Material \ref{av:A} and \ref{av:B}).

\begin{figure}[ht]
 \centering
 \includegraphics[width=\figurewidth]{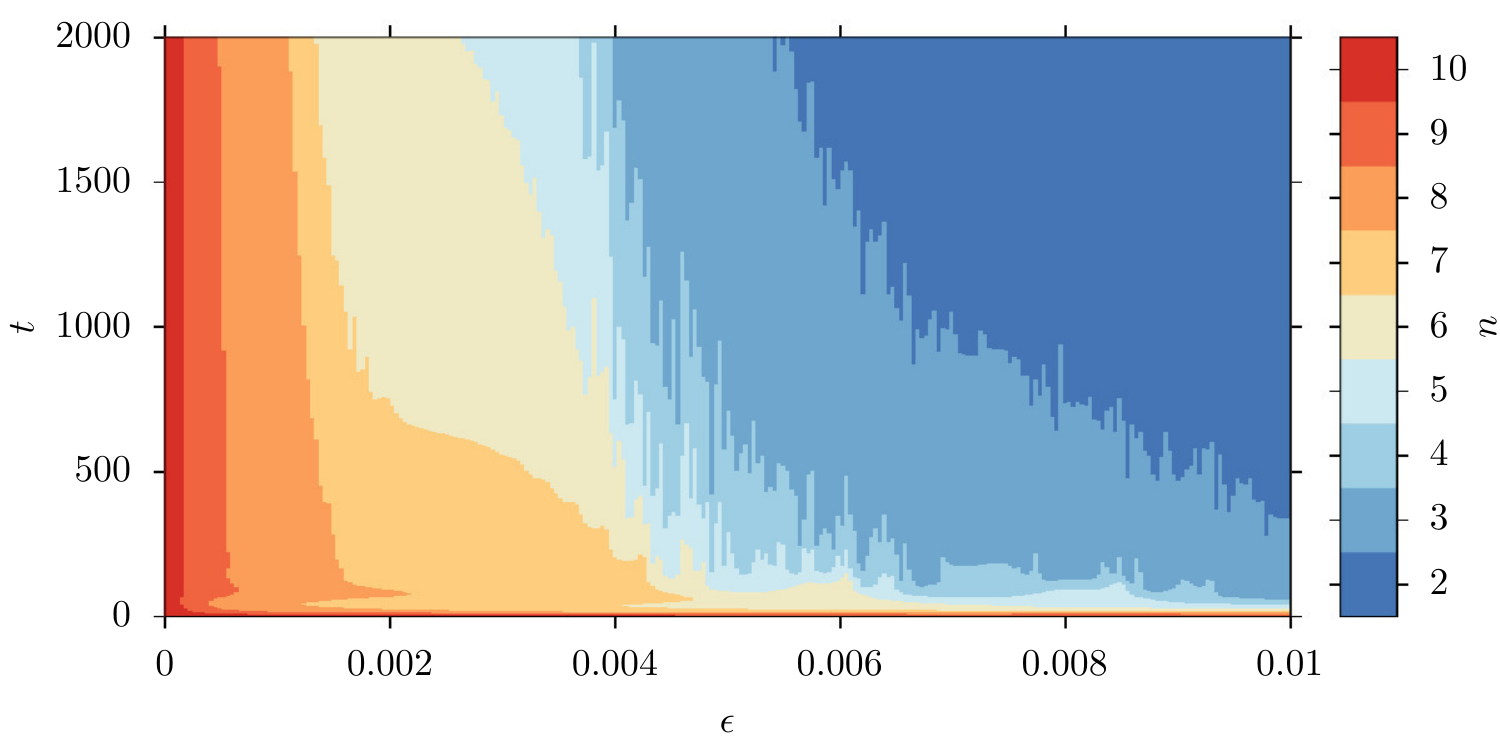}
 \caption{First radiating harmonic as a function of $\epsilon$ and evolution time starting from $\omega=0.1$  sG breather
profile.}\label{RadiationHarmonics}
\end{figure}

We have also investigated how the above patters change for different values of $\epsilon$.
In the Figure \ref{RadiationHarmonics} we have plotted the number of the first propagating harmonics as a
function of the perturbation parameter $\epsilon$ and evolution time, starting from the large amplitude sG breather
configuration with frequency $\omega=0.1$. Note that the frequency of the oscillations can be measured numerically
only when the oscillating field possess an extremum at $x=0$, which is generally not a case.
We have interpolated the frequency for much denser time gird using cubic splines.

The perturbation destroys the integrability of the model, thus even for a very small values of $\epsilon$ the
oscillon radiates the outgoing waves  reducing its amplitude and, consequently, releasing lower harmonics.
Even for relatively small value  $\epsilon\sim 0.005$,  after about $t=1000$ units of time
the oscillon relaxes radiation through the second harmonic, which becomes visible in the power spectrum.
For smaller values of $\epsilon$ the time between the transitions increases significantly.
Moreover a resonant structure is clearly visible, however the time evolution becomes much slower for some
values of $\epsilon$.

More generally, we observe that the pattern of time evolution of the radiating oscillon in the
perturbed model (\ref{potential_1}) is very sensitive to the small variations of $\epsilon$
revealing some chaotic behaviour. Note that
similar feature was observed by Gleiser for the 3+1 dimensional oscillons \cite{gleiser1994pseudostable}, it was pointed out
that some initial data produced exceptionally long living oscillons while a tiny variation of the input profile lead to
abrupt collapse of the configuration.

\begin{figure}[ht]
 \centering
 \includegraphics[width=\linewidth]{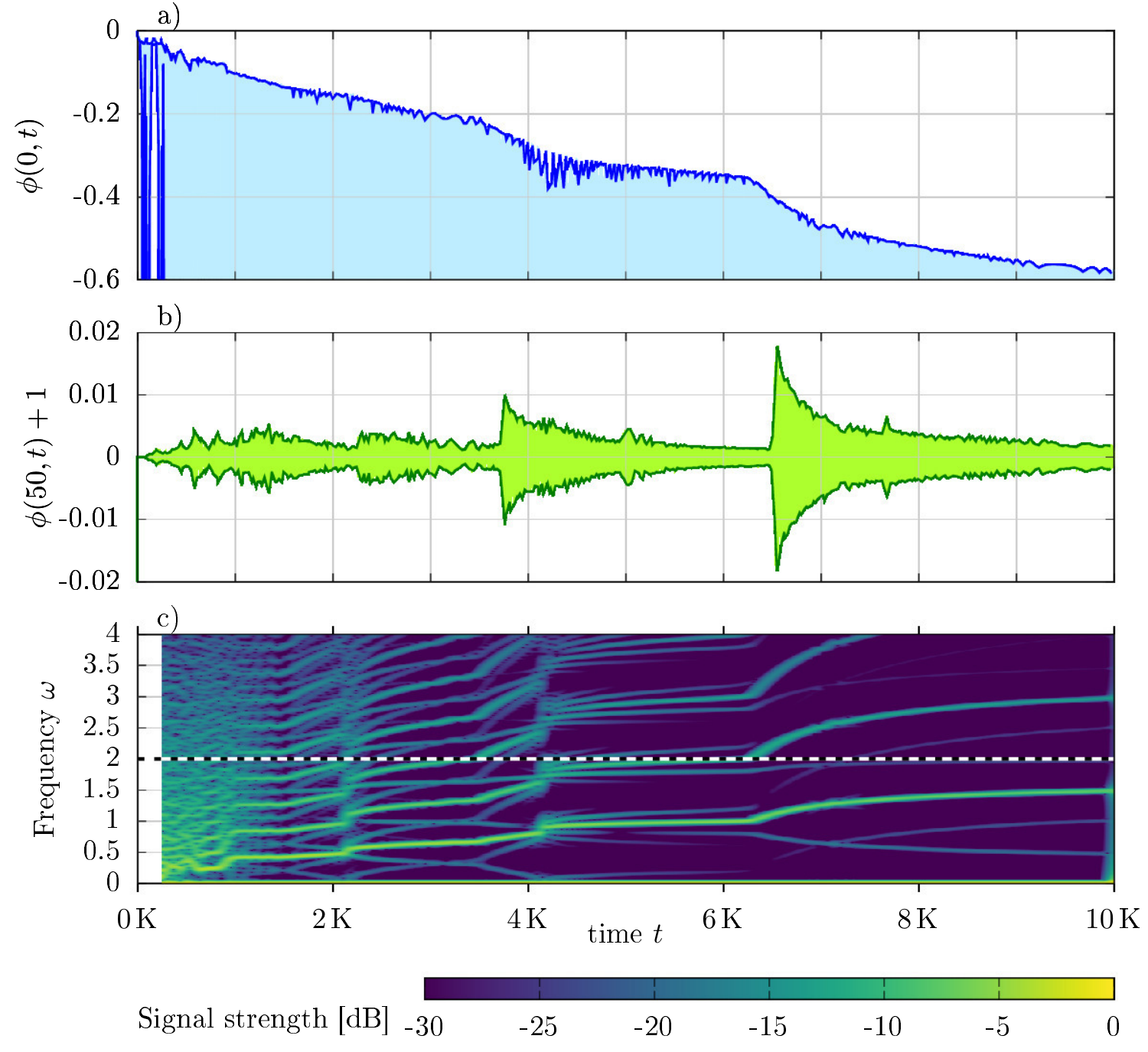}
 \caption{Multiple transitions for slow kink-antikink collisions in
a $\phi^6$ model as in \cite{Dorey:2011yw},
with initial velocity $v=0.02$. The mass threshold
$m=2$ is shown by a dashed black and white line.
a)~Amplitude of the oscillations at the centre $x=0$,
b) radiation amplitude measured at $x=50$,
c) spectrogram of the field measured at the centre  $\phi(0,t)$.}\label{phi6Evolution}
\end{figure}

\begin{figure}[ht]
 \centering
 \includegraphics[width=\linewidth]{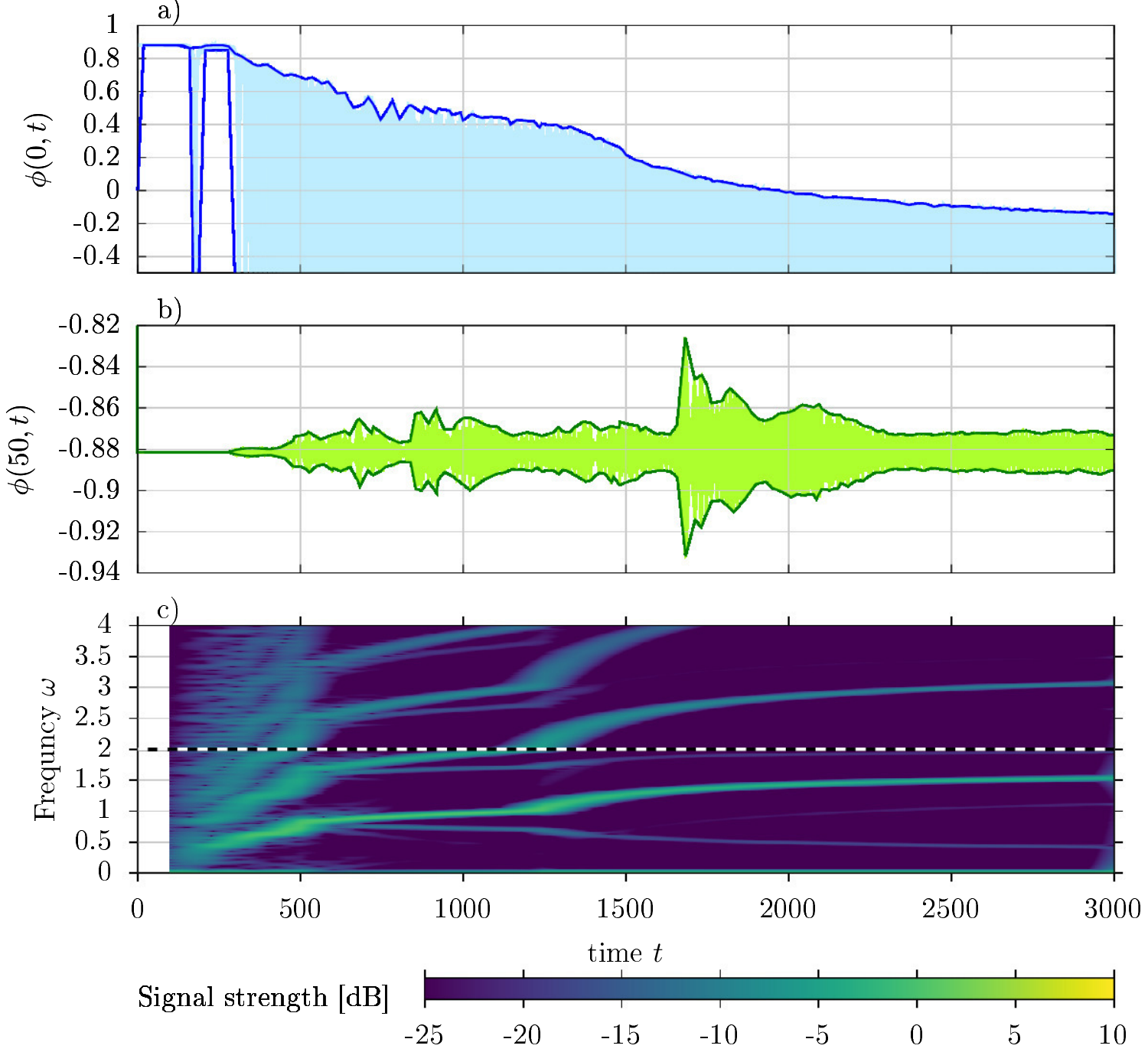}
 \caption{Transitions in the hyperbolic $\phi^4$ model after the collision of a kink-antikink pair with initial velocity $v_{in}=0.057$ a)~Amplitude of the oscillations at the centre $x=0$,
b) radiation amplitude measured at $x=50$,
c) spectrogram of the field measured at the centre  $\phi(0,t)$.}\label{hyperbolic}
\end{figure}

\noindent
\textit{~~Generalisations.~}
Further numerical simulations of our model show that for larger values
of $\epsilon$, closer to the $\phi^4$ limit, the increased coupling to
radiation means that the oscillons reduce in
amplitude more quickly,
impeding the observation of transitions 
(even for $\epsilon=0.01$, the amplitude reduces
to the point where the frequency is above $m/2$
after just $400$ units of time). However, it is not hard to find other
models where similar features are visible, including the
much-discussed $\phi^6$ model
\cite{PhysRevD.12.1606,Lohe:1979mh,Dorey:2011yw}.
In the triply-degenerate vacuum case that was the focus
of \cite{Lohe:1979mh,Dorey:2011yw}, after
a low-velocity collision of a kink and
an antikink,  a bound state (often called a
bion) is formed. This object radiates and loses energy,
slowly decaying to vacuum as an oscillon.
We have repeated the same analysis as above (Figure \ref{phi6Evolution} 
and the Supplementary Material \ref{av:C} and \ref{av:D})
for this case, looking at the collision between a right-moving
kink interpolating between the $-1$ and $0$ vacua, and a left-moving
antikink interpolating between the $0$ and $-1$ vacua,
and found that even here, relatively 
far from the sG model, the evolution
is very similar
(see figure \ref{phi6Evolution} 
and the Supplementary Material \ref{av:C} and \ref{av:D}).
The oscillon, although highly perturbed, decays
through a series of jumps which can easily be matched with the
successive
releases of higher harmonics. In the spectrogram some more
frequencies are also visible due to the initial perturbations, but
the dominant features come from the effects described in this paper.
Provided the lifting of the vacuum degeneracy is not too great,
the same phenomena can be seen in the more-general $\phi^6$ models
originally discussed in \cite{PhysRevD.12.1606}.
As a final example, we investigated
the hyperbolic $\phi^4$ model recently introduced in
\cite{Bazeia:2019xoe}, and found similar effects there too, as
shown in the Figure~\ref{hyperbolic}. 

\noindent
\textit{~~Conclusions.~}
In this paper we presented a novel approach to the analysis of
oscillons in which we treated large oscillons as perturbed breathers
in slightly deformed sG model.
As our example we took a potential which interpolates between the infinitely-degenerate and
$\mathbb{Z}$-symmetric sG potential and the doubly-degenerate potential of a scaled $\phi^4$ model.
An arbitrarily-small deformation of the sG system breaks both the integrability and the $\mathbb{Z}$
symmetry of the system, and transforms the exact time-periodic non-radiating breather
into an oscillon. The cores of these oscillating
lumps are very similar to the cores of sG breathers, but they
possess a radiating tail. One of the  surprising things we found is
that a system which is so close to an integrable model can behave in
such a complicated way (though see 
\cite{Dorey:2016bdr} for a further example of an apparently small
breaking of integrability having a dramatic effect on
a system's behaviour).
The evolution is also very different from the decay of the usual, small-amplitude oscillons, extensively studied in the literature in many models.
Counterintuitively, the large-amplitude oscillons do not decay
uniformly but rather through a series of radiative bursts of increasing
amplitude. These bursts can be associated with the release of
successive
harmonics into the continuous spectrum. Only after the releasing the
second lowest harmonic do the oscillons enter a regular decay phase.

We also discovered 
a similar type of staggered relaxation 
in the $\phi^6$ and hyperbolic $\phi^4$ models, 
in the decay of an oscillon created
by the annihilation of a kink and an antikink. 
It is natural to predict that similar phenomena will be observed in
many other models, including those
in which the integrability of the sG model
is broken 
by other field-theoretic potentials, external potentials
(or impurities),
or modified boundary conditions (including Neumann
\cite{Dorey:2017bdr} and Robin
\cite{Dorey:2016bdr}).
We therefore expect
that the effects described in this paper will arise in a wide class of field
theories, once large-amplitude excitations are considered.

Finally, we mention that non-uniform oscillon decays have
been observed in some earlier 
studies of oscillons in higher dimensions 
\cite{honda2002fine, PhysRevD.80.125037}. However, these appear to be
very different from the decays we have discussed in this
paper.
For example, the
three-dimensional oscillon decay shown in figure 6 of~\cite{honda2002fine} 
is not fully uniform, pulses of radiation initially
being released in
sync with a large period modulation of the amplitude of the field at
the origin.
However this modulation is not itself associated with the successive
crossing of thresholds, as was the case for our models.
This form of decay continues until a certain critical frequency is reached,
at which point
the oscillon of \cite{honda2002fine} becomes unstable and decays more
quickly,
releasing a large amount of radiation. 
The lifetime of this oscillon depends on the initial conditions, 
revealing an intriguing resonant structure. 
However, in these earlier studies, only oscillons with high frequency,
just below the mass threshold, were considered. 
All harmonics except the zeroth-frequency and the basic one were
already above the mass threshold. 
The observed final jump of the decay rate was due to the oscillon
entering an unstable phase, when its size was not sufficient for
nonlinearities to bind the field into a single localized object. 
We, on the other hand, studied low frequency oscillons when the jumps
in the decay rate corresponded to the sequential
 releasing of higher
 harmonics as the frequency gradually increased. Moreover, in  one
spatial dimension we would not expect an oscillon to end its
evolution with a collapse after reaching
some critical frequency, but rather to decay slowly into the
vaccum. This is indeed what we observed in our models after the final
stacatto burst of radiation was released.

\medskip

\noindent
\textit{~~Acknowledgements.~}
TR wishes to thank Piotr Bizo\'n for many inspiring discussions.
Ya.S. gratefully acknowledges partial support of the Ministry of Science and
Higher Education of Russian Federation, project No~3.1386.2017.
This project has also received funding from the European Union’s Horizon 2020 research and innovation programme under the Marie Skodowska-Curie grant agreement No. 764850, “SAGEX”, and from the STFC under consolidated grant ST/P000371/1.

\smallskip

\newcommand\movitem[1]{\item\texttt{\href{https://arxiv.org/src/1910.04128v1/anc/#1}{#1}}\,:\,}
\newcommand\movitemm[1]{\item\texttt{\href{http://th.if.uj.edu.pl/~trom/Oscillons/#1}{#1}}\,:\,}
\noindent
\textit{~~Supplementary material.~}
The files accompanying this paper
illustrate a few aspects of large oscillon decay. 
In each of the audio-visual files [A] -- [D],
there are four plots: a spectrogram (upper-left),
the field either at the centre or at $x=50$ (bottom-left), a current
FFT of a short window (upper-right) and a current view of the time
dependence of the field (bottom-right). 
We have added soundtracks to these plots,
inspired in part by the LIGO
audio files \cite{LIGO:audio}.
These soundtracks ``audibilize" the
oscillations of the field, with the frequency adjusted for
the human ear. In the far-field soundtracks the characteristic
bell-like sounds signal the passages of the staccato bursts of radiation
past the measuring point. 
\begin{enumerate}[label={[\Alph*]}]
\movitem{OscillonCentre.mov} Evolution of the field at the centre for $\epsilon=0.0025$, $\omega_0=0.1$.\label{av:A}
\movitem{OscillonFar.mov} Evolution of the field at $x=50$ for $\epsilon=0.0025$, $\omega_0=0.1$.\label{av:B}
\movitem{Phi6Centre.mov} Evolution of the field at the centre after a
kink-antikink collision  in the $\phi^6$ model.\label{av:C}
\movitem{Phi6Far.mov} Evolution of the field at $x=50$ after a
kink-antikink collision  in the $\phi^6$ model.\label{av:D}
\end{enumerate}

\bibliographystyle{elsarticle-num-names}

\bibliography{refs}

\begin{thebibliography}{33}
\expandafter\ifx\csname natexlab\endcsname\relax\def\natexlab#1{#1}\fi
\providecommand{\url}[1]{\texttt{#1}}
\providecommand{\href}[2]{#2}
\providecommand{\path}[1]{#1}
\providecommand{\DOIprefix}{doi:}
\providecommand{\ArXivprefix}{arXiv:}
\providecommand{\URLprefix}{URL: }
\providecommand{\Pubmedprefix}{pmid:}
\providecommand{\doi}[1]{\href{http://dx.doi.org/#1}{\path{#1}}}
\providecommand{\Pubmed}[1]{\href{pmid:#1}{\path{#1}}}
\providecommand{\bibinfo}[2]{#2}
\ifx\xfnm\relax \def\xfnm[#1]{\unskip,\space#1}\fi
\bibitem[{Bogolyubsky and Makhankov(1976)}]{bogolyubsky1976pulsed}
\bibinfo{author}{I.~Bogolyubsky}, \bibinfo{author}{V.~Makhankov},
\newblock \bibinfo{title}{On the pulsed soliton lifetime in two classical
  relativistic theory models},
\newblock \bibinfo{journal}{JETP Lett.} \bibinfo{volume}{24}
  (\bibinfo{year}{1976}) \bibinfo{pages}{12}.
\bibitem[{Copeland et~al.(1995)Copeland, Gleiser, and
  M{\"u}ller}]{copeland1995oscillons}
\bibinfo{author}{E.~J. Copeland}, \bibinfo{author}{M.~Gleiser},
  \bibinfo{author}{H.-R. M{\"u}ller},
\newblock \bibinfo{title}{Oscillons: Resonant configurations during bubble
  collapse},
\newblock \bibinfo{journal}{Phys. Rev. D} \bibinfo{volume}{52}
  (\bibinfo{year}{1995}) \bibinfo{pages}{1920}. \URLprefix
  \url{https://doi.org/10.1103/PhysRevD.52.1920}.
\bibitem[{Gleiser(1994)}]{gleiser1994pseudostable}
\bibinfo{author}{M.~Gleiser},
\newblock \bibinfo{title}{Pseudostable bubbles},
\newblock \bibinfo{journal}{Phys. Rev. D} \bibinfo{volume}{49}
  (\bibinfo{year}{1994}) \bibinfo{pages}{2978}. \URLprefix
  \url{https://doi.org/10.1103/PhysRevD.49.2978}.
\bibitem[{Riotto(1996)}]{Riotto:1995yy}
\bibinfo{author}{A.~Riotto},
\newblock \bibinfo{title}{{Are oscillons present during a first order
  electroweak phase transition?}},
\newblock \bibinfo{journal}{Phys. Lett.} \bibinfo{volume}{B365}
  (\bibinfo{year}{1996}) \bibinfo{pages}{64--71}.
  \DOIprefix\doi{10.1016/0370-2693(95)01239-7}.
  \href{http://arxiv.org/abs/hep-ph/9507201}{{\tt arXiv:hep-ph/9507201}}.
\bibitem[{Umurhan et~al.(1998)Umurhan, Tao, and Spiegel}]{Umurhan:1998ez}
\bibinfo{author}{O.~M. Umurhan}, \bibinfo{author}{L.~Tao},
  \bibinfo{author}{E.~A. Spiegel},
\newblock \bibinfo{title}{{Stellar oscillons}},
\newblock \bibinfo{journal}{Annals N. Y. Acad. Sci.} \bibinfo{volume}{867}
  (\bibinfo{year}{1998}) \bibinfo{pages}{298}.
  \DOIprefix\doi{10.1111/j.1749-6632.1998.tb11265.x}.
  \href{http://arxiv.org/abs/astro-ph/9806209}{{\tt arXiv:astro-ph/9806209}}.
\bibitem[{Farhi et~al.(2005)Farhi, Graham, Khemani, Markov, and
  Rosales}]{Farhi:2005rz}
\bibinfo{author}{E.~Farhi}, \bibinfo{author}{N.~Graham},
  \bibinfo{author}{V.~Khemani}, \bibinfo{author}{R.~Markov},
  \bibinfo{author}{R.~Rosales},
\newblock \bibinfo{title}{{An Oscillon in the SU(2) gauged Higgs model}},
\newblock \bibinfo{journal}{Phys. Rev.} \bibinfo{volume}{D72}
  (\bibinfo{year}{2005}) \bibinfo{pages}{101701}.
  \DOIprefix\doi{10.1103/PhysRevD.72.101701}.
  \href{http://arxiv.org/abs/hep-th/0505273}{{\tt arXiv:hep-th/0505273}}.
\bibitem[{Gleiser(2007)}]{Gleiser:2006te}
\bibinfo{author}{M.~Gleiser},
\newblock \bibinfo{title}{{Oscillons in scalar field theories: Applications in
  higher dimensions and inflation}},
\newblock \bibinfo{journal}{Int. J. Mod. Phys.} \bibinfo{volume}{D16}
  (\bibinfo{year}{2007}) \bibinfo{pages}{219--229}.
  \DOIprefix\doi{10.1142/S0218271807009954}.
  \href{http://arxiv.org/abs/hep-th/0602187}{{\tt arXiv:hep-th/0602187}}.
\bibitem[{Alcubierre et~al.(2003)Alcubierre, Becerril, Guzman, Matos, Nunez,
  and Urena-Lopez}]{Alcubierre:2003sx}
\bibinfo{author}{M.~Alcubierre}, \bibinfo{author}{R.~Becerril},
  \bibinfo{author}{S.~F. Guzman}, \bibinfo{author}{T.~Matos},
  \bibinfo{author}{D.~Nunez}, \bibinfo{author}{L.~A. Urena-Lopez},
\newblock \bibinfo{title}{{Numerical studies of $\Phi^2$-oscillatons}},
\newblock \bibinfo{journal}{Class. Quant. Grav.} \bibinfo{volume}{20}
  (\bibinfo{year}{2003}) \bibinfo{pages}{2883--2904}.
  \DOIprefix\doi{10.1088/0264-9381/20/13/332}.
  \href{http://arxiv.org/abs/gr-qc/0301105}{{\tt arXiv:gr-qc/0301105}}.
\bibitem[{Graham(2007)}]{Graham:2006vy}
\bibinfo{author}{N.~Graham},
\newblock \bibinfo{title}{{An Electroweak oscillon}},
\newblock \bibinfo{journal}{Phys. Rev. Lett.} \bibinfo{volume}{98}
  (\bibinfo{year}{2007}) \bibinfo{pages}{101801}.
  \DOIprefix\doi{10.1103/PhysRevLett.98.101801, 10.1103/PhysRevLett.98.189904}.
  \href{http://arxiv.org/abs/hep-th/0610267}{{\tt arXiv:hep-th/0610267}},
  \bibinfo{note}{[Erratum: Phys. Rev. Lett. {\bf 98}, 189904 (2007)]}.
\bibitem[{Fodor et~al.(2010)Fodor, Forgacs, and Mezei}]{Fodor:2009kg}
\bibinfo{author}{G.~Fodor}, \bibinfo{author}{P.~Forgacs},
  \bibinfo{author}{M.~Mezei},
\newblock \bibinfo{title}{{Mass loss and longevity of gravitationally bound
  oscillating scalar lumps (oscillatons) in D-dimensions}},
\newblock \bibinfo{journal}{Phys. Rev.} \bibinfo{volume}{D81}
  (\bibinfo{year}{2010}) \bibinfo{pages}{064029}.
  \DOIprefix\doi{10.1103/PhysRevD.81.064029}.
  \href{http://arxiv.org/abs/0912.5351}{{\tt arXiv:0912.5351}}.
\bibitem[{Cuevas et~al.(2009)Cuevas, English, Kevrekidis, and
  Anderson}]{PhysRevLett.102.224101}
\bibinfo{author}{J.~Cuevas}, \bibinfo{author}{L.~Q. English},
  \bibinfo{author}{P.~G. Kevrekidis}, \bibinfo{author}{M.~Anderson},
\newblock \bibinfo{title}{Discrete breathers in a forced-damped array of
  coupled pendula: Modeling, computation, and experiment},
\newblock \bibinfo{journal}{Phys. Rev. Lett.} \bibinfo{volume}{102}
  (\bibinfo{year}{2009}) \bibinfo{pages}{224101}. \URLprefix
  \url{https://link.aps.org/doi/10.1103/PhysRevLett.102.224101}.
  \DOIprefix\doi{10.1103/PhysRevLett.102.224101}.
\bibitem[{Kivshar and Malomed(1989)}]{RevModPhys.61.763}
\bibinfo{author}{Y.~S. Kivshar}, \bibinfo{author}{B.~A. Malomed},
\newblock \bibinfo{title}{Dynamics of solitons in nearly integrable systems},
\newblock \bibinfo{journal}{Rev. Mod. Phys.} \bibinfo{volume}{61}
  (\bibinfo{year}{1989}) \bibinfo{pages}{763--915}. \URLprefix
  \url{https://link.aps.org/doi/10.1103/RevModPhys.61.763}.
  \DOIprefix\doi{10.1103/RevModPhys.61.763}.
\bibitem[{Cuevas-Maraver and Kevrekidis(2019)}]{Cuevas–Maraver2019}
\bibinfo{author}{J.~Cuevas-Maraver}, \bibinfo{author}{P.~G. Kevrekidis},
  \bibinfo{title}{Discrete Breathers in $\phi^4$ and Related Models},
  \bibinfo{publisher}{Springer International Publishing},
  \bibinfo{address}{Cham}, \bibinfo{year}{2019}, pp. \bibinfo{pages}{137--162}.
  \URLprefix \url{https://doi.org/10.1007/978-3-030-11839-6_7}.
  \DOIprefix\doi{10.1007/978-3-030-11839-6_7}.
\bibitem[{Cuevas-Maraver et~al.(2014)Cuevas-Maraver, Kevrekidis, and
  Williams}]{sgBook2014}
\bibinfo{editor}{J.~Cuevas-Maraver}, \bibinfo{editor}{P.~G. Kevrekidis},
  \bibinfo{editor}{F.~Williams} (Eds.), \bibinfo{title}{The sine-Gordon Model
  and its Applications: From Pendula and Josephson Junctions to Gravity and
  High-Energy Physics}, \bibinfo{publisher}{Springer International Publishing},
  \bibinfo{address}{Cham}, \bibinfo{year}{2014}, pp. \bibinfo{pages}{1--30}.
\bibitem[{Gulevich et~al.(2012)Gulevich, Gaifullin, and Kusmartsev}]{Gulevich}
\bibinfo{author}{D.~R. Gulevich}, \bibinfo{author}{M.~B. Gaifullin},
  \bibinfo{author}{F.~V. Kusmartsev},
\newblock \bibinfo{title}{Controlled dynamics of sine-gordon breather in long
  josephson junctions},
\newblock \bibinfo{journal}{The European Physical Journal B}
  \bibinfo{volume}{85} (\bibinfo{year}{2012}) \bibinfo{pages}{24}. \URLprefix
  \url{https://doi.org/10.1140/epjb/e2011-20689-4}.
  \DOIprefix\doi{10.1140/epjb/e2011-20689-4}.
\bibitem[{Flach and Willis(1998)}]{FLACH1998181}
\bibinfo{author}{S.~Flach}, \bibinfo{author}{C.~Willis},
\newblock \bibinfo{title}{Discrete breathers},
\newblock \bibinfo{journal}{Physics Reports} \bibinfo{volume}{295}
  (\bibinfo{year}{1998}) \bibinfo{pages}{181 -- 264}. \URLprefix
  \url{http://www.sciencedirect.com/science/article/pii/S0370157397000689}.
  \DOIprefix\doi{https://doi.org/10.1016/S0370-1573(97)00068-9}.
\bibitem[{Christ and Lee(1975)}]{PhysRevD.12.1606}
\bibinfo{author}{N.~H. Christ}, \bibinfo{author}{T.~D. Lee},
\newblock \bibinfo{title}{Quantum expansion of soliton solutions},
\newblock \bibinfo{journal}{Phys. Rev. D} \bibinfo{volume}{12}
  (\bibinfo{year}{1975}) \bibinfo{pages}{1606--1627}. \URLprefix
  \url{https://link.aps.org/doi/10.1103/PhysRevD.12.1606}.
  \DOIprefix\doi{10.1103/PhysRevD.12.1606}.
\bibitem[{Lohe(1979)}]{Lohe:1979mh}
\bibinfo{author}{M.~A. Lohe},
\newblock \bibinfo{title}{Soliton structures in
  $p{(\ensuremath{\varphi})}_{2}$},
\newblock \bibinfo{journal}{Phys. Rev. D} \bibinfo{volume}{20}
  (\bibinfo{year}{1979}) \bibinfo{pages}{3120--3130}. \URLprefix
  \url{https://link.aps.org/doi/10.1103/PhysRevD.20.3120}.
  \DOIprefix\doi{10.1103/PhysRevD.20.3120}.
\bibitem[{Segur and Kruskal(1987)}]{segur1987nonexistence}
\bibinfo{author}{H.~Segur}, \bibinfo{author}{M.~D. Kruskal},
\newblock \bibinfo{title}{{Nonexistence of small-amplitude breather solutions
  in $\phi^4$ theory}},
\newblock \bibinfo{journal}{Phys. Rev. Lett.} \bibinfo{volume}{58}
  (\bibinfo{year}{1987}) \bibinfo{pages}{747}. \URLprefix
  \url{https://doi.org/10.1103/PhysRevLett.58.747}.
\bibitem[{Boyd(1995)}]{BOYD1995311}
\bibinfo{author}{J.~P. Boyd},
\newblock \bibinfo{title}{{Weakly nonlocal envelope solitary waves: numerical
  calculations for the Klein-Gordon ($\phi^4$) equation}},
\newblock \bibinfo{journal}{Wave Motion} \bibinfo{volume}{21}
  (\bibinfo{year}{1995}) \bibinfo{pages}{311 -- 330}. \URLprefix
  \url{http://www.sciencedirect.com/science/article/pii/0165212595000054}.
  \DOIprefix\doi{https://doi.org/10.1016/0165-2125(95)00005-4}.
\bibitem[{Fodor et~al.(2009)Fodor, Forgacs, Horv{\'a}th, and
  Mezei}]{fodor2009computation}
\bibinfo{author}{G.~Fodor}, \bibinfo{author}{P.~Forgacs},
  \bibinfo{author}{Z.~Horv{\'a}th}, \bibinfo{author}{M.~Mezei},
\newblock \bibinfo{title}{Computation of the radiation amplitude of oscillons},
\newblock \bibinfo{journal}{Phys. Rev. D} \bibinfo{volume}{79}
  (\bibinfo{year}{2009}) \bibinfo{pages}{065002}. \URLprefix
  \url{https://doi.org/10.1103/PhysRevD.79.065002}.
\bibitem[{Fodor(2019)}]{Fodor:2019ftc}
\bibinfo{author}{G.~Fodor}, \bibinfo{title}{{A review on radiation of oscillons
  and oscillatons}}, Ph.D. thesis, Wigner RCP, Budapest, \bibinfo{year}{2019}.
  \href{http://arxiv.org/abs/1911.03340}{{\tt arXiv:1911.03340}}.
\bibitem[{Hindmarsh and Salmi(2008)}]{hindmarsh2008oscillons}
\bibinfo{author}{M.~Hindmarsh}, \bibinfo{author}{P.~Salmi},
\newblock \bibinfo{title}{Oscillons and domain walls},
\newblock \bibinfo{journal}{Phys. Rev. D} \bibinfo{volume}{77}
  (\bibinfo{year}{2008}) \bibinfo{pages}{105025}. \URLprefix
  \url{https://doi.org/10.1103/PhysRevD.77.105025}.
\bibitem[{Honda and Choptuik(2002)}]{honda2002fine}
\bibinfo{author}{E.~P. Honda}, \bibinfo{author}{M.~W. Choptuik},
\newblock \bibinfo{title}{{Fine structure of oscillons in the spherically
  symmetric $\phi^4$ Klein-Gordon model}},
\newblock \bibinfo{journal}{Phys. Rev. D} \bibinfo{volume}{65}
  (\bibinfo{year}{2002}) \bibinfo{pages}{084037}.
\bibitem[{Salmi and Hindmarsh(2012)}]{Hindmarsh2012}
\bibinfo{author}{P.~Salmi}, \bibinfo{author}{M.~Hindmarsh},
\newblock \bibinfo{title}{Radiation and relaxation of oscillons},
\newblock \bibinfo{journal}{Phys. Rev. D} \bibinfo{volume}{85}
  (\bibinfo{year}{2012}) \bibinfo{pages}{085033}. \URLprefix
  \url{https://link.aps.org/doi/10.1103/PhysRevD.85.085033}.
  \DOIprefix\doi{10.1103/PhysRevD.85.085033}.
  \href{http://arxiv.org/abs/1201.1934}{{\tt arXiv:1201.1934}}.
\bibitem[{Gleiser and Sicilia(2008)}]{PhysRevLett.101.011602}
\bibinfo{author}{M.~Gleiser}, \bibinfo{author}{D.~Sicilia},
\newblock \bibinfo{title}{Analytical characterization of oscillon energy and
  lifetime},
\newblock \bibinfo{journal}{Phys. Rev. Lett.} \bibinfo{volume}{101}
  (\bibinfo{year}{2008}) \bibinfo{pages}{011602}. \URLprefix
  \url{https://link.aps.org/doi/10.1103/PhysRevLett.101.011602}.
  \DOIprefix\doi{10.1103/PhysRevLett.101.011602}.
\bibitem[{Gleiser and Sicilia(2009)}]{PhysRevD.80.125037}
\bibinfo{author}{M.~Gleiser}, \bibinfo{author}{D.~Sicilia},
\newblock \bibinfo{title}{General theory of oscillon dynamics},
\newblock \bibinfo{journal}{Phys. Rev. D} \bibinfo{volume}{80}
  (\bibinfo{year}{2009}) \bibinfo{pages}{125037}. \URLprefix
  \url{https://link.aps.org/doi/10.1103/PhysRevD.80.125037}.
  \DOIprefix\doi{10.1103/PhysRevD.80.125037}.
\bibitem[{Dorey et~al.(2017)Dorey, Halavanau, Mercer, Roma\'nczukiewicz, and
  Shnir}]{Dorey:2017bdr}
\bibinfo{author}{P.~Dorey}, \bibinfo{author}{A.~Halavanau},
  \bibinfo{author}{J.~Mercer}, \bibinfo{author}{T.~Roma\'nczukiewicz},
  \bibinfo{author}{Y.~Shnir},
\newblock \bibinfo{title}{{Boundary scattering in the $\phi^{4}$ model}},
\newblock \bibinfo{journal}{JHEP} \bibinfo{volume}{2017} (\bibinfo{year}{2017})
  \bibinfo{pages}{107}. \DOIprefix\doi{10.1007/JHEP05(2017)107}.
  \href{http://arxiv.org/abs/1508.02329}{{\tt arXiv:1508.02329}}.
\bibitem[{Roma\'nczukiewicz and Shnir(2018)}]{OscillonsExternal2018}
\bibinfo{author}{T.~Roma\'nczukiewicz}, \bibinfo{author}{Y.~Shnir},
\newblock \bibinfo{title}{{Oscillons in the presence of external potential}},
\newblock \bibinfo{journal}{JHEP} \bibinfo{volume}{2018} (\bibinfo{year}{2018})
  \bibinfo{pages}{101}. \URLprefix
  \url{https://doi.org/10.1007/JHEP01(2018)101}.
  \DOIprefix\doi{10.1007/JHEP01(2018)101}.
  \href{http://arxiv.org/abs/1706.09234}{{\tt arXiv:1706.09234}}.
\bibitem[{Dorey et~al.(2011)Dorey, Mersh, Roma\'nczukiewicz, and
  Shnir}]{Dorey:2011yw}
\bibinfo{author}{P.~Dorey}, \bibinfo{author}{K.~Mersh},
  \bibinfo{author}{T.~Roma\'nczukiewicz}, \bibinfo{author}{{\relax Ya}.~Shnir},
\newblock \bibinfo{title}{{Kink-antikink collisions in the $\phi^6$ model}},
\newblock \bibinfo{journal}{Phys. Rev. Lett.} \bibinfo{volume}{107}
  (\bibinfo{year}{2011}) \bibinfo{pages}{091602}.
  \DOIprefix\doi{10.1103/PhysRevLett.107.091602}.
  \href{http://arxiv.org/abs/1101.5951}{{\tt arXiv:1101.5951}}.
\bibitem[{Bazeia et~al.(2019)Bazeia, Gomes, Nobrega, and
  Simas}]{Bazeia:2019xoe}
\bibinfo{author}{D.~Bazeia}, \bibinfo{author}{A.~R. Gomes},
  \bibinfo{author}{K.~Z. Nobrega}, \bibinfo{author}{F.~C. Simas},
\newblock \bibinfo{title}{Kink scattering in hyperbolic models},
\newblock \bibinfo{journal}{International Journal of Modern Physics A}
  \bibinfo{volume}{34} (\bibinfo{year}{2019}) \bibinfo{pages}{1950200}.
  \URLprefix \url{https://doi.org/10.1142/S0217751X19502002}.
  \DOIprefix\doi{10.1142/S0217751X19502002}.
  \href{http://arxiv.org/abs/1902.04041}{{\tt arXiv:1902.04041}}.
\bibitem[{Arthur et~al.(2016)Arthur, Dorey, and Parini}]{Dorey:2016bdr}
\bibinfo{author}{R.~Arthur}, \bibinfo{author}{P.~Dorey},
  \bibinfo{author}{R.~Parini},
\newblock \bibinfo{title}{{Breaking integrability at the boundary : the
  sine-Gordon model with Robin boundary conditions}},
\newblock \bibinfo{journal}{J. Phys. A} \bibinfo{volume}{49}
  (\bibinfo{year}{2016}) \bibinfo{pages}{165205}. \URLprefix
  \url{https://dx.doi.org/10.1088/1751-8113/49/16/165205}.
  \href{http://arxiv.org/abs/1509.08448}{{\tt arXiv:1509.08448}}.
\bibitem[{{LIGO Scientific Collaboration and Virgo
  Collaboration}(2019)}]{LIGO:audio}
\bibinfo{author}{{LIGO Scientific Collaboration and Virgo Collaboration}},
  \bibinfo{title}{Gravitational wave open science center: Audio files},
  \bibinfo{howpublished}{\url{https://www.gw-openscience.org/audio/}},
  \bibinfo{year}{access: 2019}.

\end{thebibliography}
\end{document}